\documentclass[authoryear,preprint,review,10pt]{elsarticle}

\usepackage{graphicx}
\usepackage{float}
\usepackage{amssymb}
\usepackage{rotfloat}

\journal{Astroparticle Physics}

\begin{document}
\begin{frontmatter}
\title{Toward an understanding of thermal X-ray emission of pulsars}

\author{M. Yu}
\author{R. X. Xu\corref{xu}}

\ead{r.x.xu@pku.edu.cn}

\cortext[xu]{Tel.: +86 10 62758631; fax: +86 10 62765031}

\address{School of Physics and State Key Laboratory of Nuclear Physics and
Technology, Peking University, Beijing 100871, China}
\begin{abstract}
We present a theoretical model for the thermal X-ray emission properties and cooling behaviors of isolated
pulsars, assuming that pulsars are solid quark stars.
We calculate the heat capacity for such a quark star, including the component of the crystalline lattice and
that of the extremely relativistic electron gas. The results show that the residual thermal energy cannot
sustain the observed thermal X-ray luminosities seen in typical isolated X-ray pulsars.
We conclude that other heating mechanisms must be in operation if the pulsars are in fact solid quark stars. Two
possible heating mechanisms are explored. Firstly, for pulsars with little magnetospheric activities, accretion
from the interstellar medium or from the material in the associated supernova remnants may power the observed
thermal emission. In the propeller regime, a disk-accretion rate ${\dot M}\sim$1 \% of the Eddington rate with
an accretion onto the stellar surface at a rate of $\sim 0.1\% {\dot M}$ could explain the observed emission
luminosities of the dim isolated neutron stars and the central compact objects. Secondly, for pulsars with
significant magnetospheric activities, the pulsar spindown luminosities may have been as the sources of the
thermal energy via reversing plasma current flows. A phenomenological study between pulsar bolometric X-ray
luminosities and the spin energy loss rates presents the probable existence of a 1/2-law or a linear law, i.e.
$L_{\rm bol}^{\infty}\propto\dot{E}^{1/2}$ or $L_{\rm bol}^{\infty}\propto\dot{E}$. This result together with
the thermal properties of solid quark stars allow us to calculate the thermal evolution of such stars. Thermal
evolution curves, or cooling curves, are calculated and compared with the `temperature-age' data obtained from
17 active X-ray pulsars. It is shown that the bolometric X-ray observations of these sources are consistent with
the solid quark star pulsar model.

\vspace{5mm} \noindent {\it PACS:}
97.60.Gb, 97.60.Jd, 95.30.Cq%
\end{abstract}
\begin{keyword}
Pulsars; Neutron stars; Elementary particles
\end{keyword}
\end{frontmatter}
\section{Introduction}
It is conventionally thought that the thermal X-ray components of neutron stars are originated from the initial
residual heat when the stars cool~\citep{Tsuruta09}, even before the discovery of Galactic X-ray sources and
pulsars.
However, we will focus on this old problem in the regime of quark stars since there is no clear observational
evidence to rule out quark stars or neutron stars.
We demonstrate in this paper that the observed thermal emission of isolated pulsars could be well understood in
a solid quark star model.

The study of quark matter phases, both hot and cold, has been an interesting topic of research in recent years.
In an astrophysical context, quark stars composed by cold quark matter have yet not been ruled out by the
measured properties of pulsar-like compact stars~\citep{xu09}.
It has recently been proposed that realistic quark matter in compact stars could exist in a solid state
\citep{xu03,Horvath05,Owen05,mrs07}, either as a super-solid or as a normal solid~\citep{xu09}.
The basic conjecture of normal solid quark matter is that
de-confined quarks tend to form quark-clusters when the temperature
and density are relatively low. Below a certain critical
temperature, these clusters could be in periodic lattices immersed
in a degenerate, extremely relativistic, electron gas. Note that
even though quark matter is usually described as weakly coupled, the
interaction between quarks and gluons in a quark-gluon plasma is
still very strong \citep{Shuryak}. It is this strong coupling that
could cause the quarks to cluster and form a solid-like material.

In fact, various kinds of observations could put constraints on the state of the matter in a pulsar, and the cooling
behavior is suggested since 1960s, even before the discovery of pulsars. Can contemporary observations of X-ray
thermal emitting and cooling pulsars be understood in the proposed solid quark star (hereafter SQS) model? This
is a question we try to answer in the paper, and we address that SQSs could not be ruled out by the thermal
observations. Moreover, SQSs might provide an opportunity to evaluate pulsar moments of inertia, if the
thermal emission is also powered by spin.

In this paper, we argue that contemporary observations of thermal X-ray emitting pulsars are consistent with the
assumption that these sources are in fact SQSs.
The present analysis and calculation will concentrate on the
observational temperature range, i.e. a few hundreds of to a few
tens of eV. Nonetheless, a phenomenological scenario for the whole
thermal history of a strange quark star is also outlined, as
described in \S2.1.
The stellar residual heat would be the first energy source supporting the X-ray bolometric luminosities of SQSs.
\S2.2 is involved on this topic, including both of the partial contributions of the lattices and the electrons.
Other energy reservoirs could exist as the stellar heating processes. This will be discussed in \S2.3, where
different heating mechanisms are introduced, according to the different X-ray pulsar manifestations.
In \S3, we compare the observations and the predictions given by the
SQS pulsar model. Our conclusions as well as further discussions are
presented in \S4.
\section{The model}
\subsection{Cooling stages of a quark star}
\begin{figure}
\begin{center}
\includegraphics[width=0.8\textwidth]{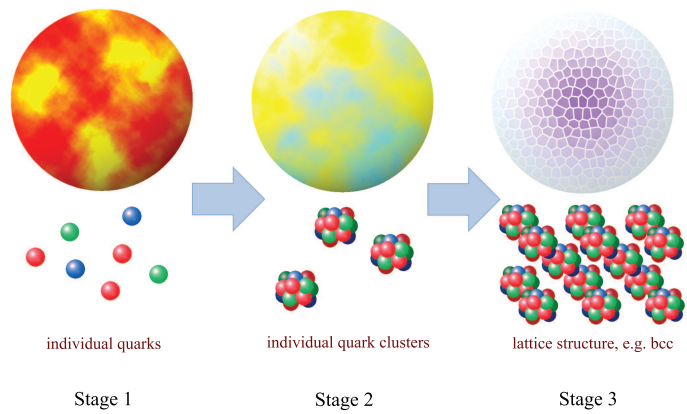}
\caption{Possible cooling stages of a quark star.
{\em Stage 1}: individual quark phase. The temperature is high ($> 10$ MeV) when the star is born, and the state
of the star could be the fluid of individual quarks.
{\em Stage 2}: individual quark cluster phase. As the temperature decreases, individual quarks tend to form
quark clusters because of the strong coupling between them. The state of the star could then be the fluid of
quark clusters.
{\em Stage 3}: solid quark star phase. As the temperature drops to a
melting temperature, the fluid of quark clusters tend to solidify to
form periodic lattice structure, such as bcc
structure.}\label{fig:stg}
\end{center}
\end{figure}
The cooling process of a quark star can be quite complicated to model when starting from the birth of the star.
Theoretical uncertainties make it difficult to predict the exact temperatures where phase transitions occur. For
illustrative purposes, we present the following scenario where cooling of a quark star takes place in
approximately three stages (See Fig. \ref{fig:stg}). The first stage occurs just after the birth of the quark
star if its initial temperature is much higher than $\sim 10^{11}$ K (10 MeV).  The emission of neutrinos and
photons will lead to fast cooling. Hence, the star quickly enters the second stage where de-confined quarks
begin to form quark clusters. As the temperature drops further, the fluid solidifies and the star is said to
enter the third cooling stage or the SQS phase. Here, quark clusters form periodic lattice structures (e.g. the
bcc structure).

It is uncertain exactly how long the star will spend in each phase, or even if it will ever enter all three
states. If the initial temperature is just around $10^{11}$ K or lower, Stage 1 may be short lived or even
non-existent. The quark-gluon plasma could be strongly coupled at birth and quark-clusters could be present
immediately after formation. It is even possible that the melting temperature of solid quark matter could be
$\sim10^{11}$ K or higher. Hence, a quark star would enter Stage 3 immediately.

The mechanisms for the emission of thermal photons and neutrinos in each specific stage would be quite
different.
A hot quark star (in Stages 1 or 2) would be a good radiator for thermal equilibrium photons with energy more
than $\sim20$ MeV \citep{alc86}. Meanwhile, the intense release of thermal energy would stimulate the generation
and radiation of electron-positron pairs from the hot bare quark surface. The annihilation of the
electron-positron pairs would generate the emission of photons inversely, and this plasma could be optically
thick enough to produce a black body spectrum \citep{usov01}. The emission of neutrinos would be given rise to
by the pair production process as well as the plasma process induced by the ultra-relativistic degenerate
electron gas inside the star \citep{it89}. The basic URCA process would be another neutrino radiation component
for Stage 1.
When the star cools down and becomes solidified (in Stage 3),
thermal equilibrium photons (usually soft X-ray) could also be
emitted from the bare SQS surface as free electrons transit in the
levels of the energy bands of solid quark matter \citep{xu03}. The
electron-phonon interaction as well as the interaction between
electrons themselves might result in a metal-like spectrum.
\citet{zxz04} analyzed the spectrum of RX J1856 phenomenologically,
and they did not conclude that there are significant differences
between the metal-like spectrum and black body spectrum. Hence, a
black body thermal spectrum would be a good approximation for SQSs.
The neutrino emissivity of clustered quark matter is theoretically so far unknown. The pair production and the
plasma process would also lead the neutrino cooling for SQSs, if the stars enter the Stage 3 at high
temperatures, such as $\sim10^{11}$ K. Nevertheless, in the observational low temperature range $\sim10^6-10^5$
K, the neutrino luminosity would be low enough so that photon cooling would be the dominant process for SQSs.

In general, in the low temperature range ($\sim10^6-10^5$ K) which is the case considered by the present paper,
the SQSs cool down by losing thermal photons, and the sources of the energy that enable the thermal emission are
firstly the residual thermal energy of the stars and secondly the energy input caused by the processes of
stellar heating.
The energy relation would then be described as
\begin{equation}
-\frac{{\rm d}U}{{\rm d}t}+L_{\rm SH}= L_{\rm bol}, \label{eq:coolinggeneral}
\end{equation}
where $L_{\rm bol}$ is the stellar bolometric X-ray luminosity, $-{\rm d}U/{\rm d}t$ is the release rate of the
stellar residual thermal energy, and $L_{\rm SH}$ is the luminosity of stellar heating.
\subsection{Residual thermal energy of a SQS}
\begin{figure}
\begin{center}
\includegraphics[width=0.8\textwidth]{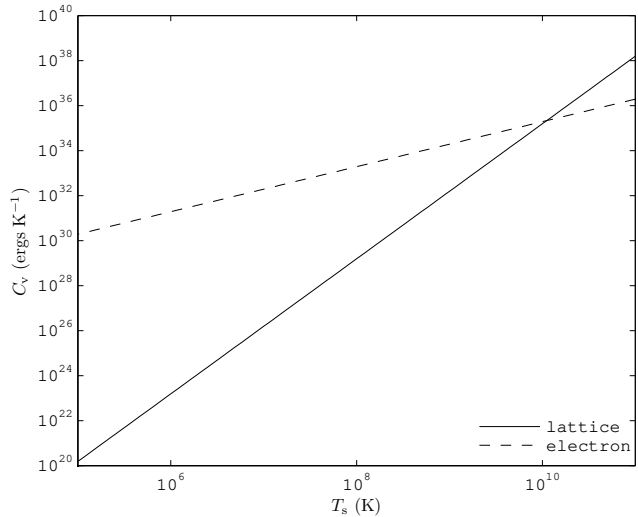}
\caption{Heat capacities of the lattice and electron components. The
solid line is the result of the lattice structure, and the dashed
line is that of the electron gas.}\label{fig:cv}
\end{center}
\end{figure}
Provided that the volume of a SQS is a constant, then the stellar residual thermal energy $U_{\rm SQS}$ would
only be the function of the stellar temperature $T_{\rm s}$,
\begin{equation}
U_{\rm SQS}(T_{\rm s})=\int C_{v}{\rm d}T_{\rm s}, \label{eq:inner}
\end{equation}
where $T_{\rm s}$ is the value in the star's local reference frame. The heat capacity of the star $C_v$
comprises of the partial contribution of the lattice structure $C_v^{\rm l}$ and that of the degenerate
electrons $C_v^{\rm e}$, or $C_v=C_v^{\rm l}+C_v^{\rm e}$.

Following Debye elastic medium theory, the characteristic of the lattice heat capacity of solid quark matter
could be evaluated by Debye temperature,
\begin{equation}
\theta_{D}=\hbar \omega_D/k_{\rm B}, \label{eq:debyeT}
\end{equation}
where $\hbar$ is the reduced Planck constant, and $k_{\rm B}$ is the Boltzmann constant.
$\omega_D$ is Debye cut-off frequency (i.e. the maximum frequency of the wave that could propagate in a medium),
which equals Debye wave number $k_{D}=(6\pi^2n_{\rm c})^{1/3}$($n_{\rm c}$ is the number density of classical
particles, or quark clusters for solid quark matter) times the average sound speed in the medium, i.e.
$\omega_D=k_{D}\bar{c}_s$.
For a SQS, the average sound speed could be the light speed approximately. A linear equation of state, extended
to be used for a quark star, indicates that the pressure $p\sim \rho c^2$, where $\rho$ is the mass density of a
quark star and $c$ is the speed of light. So an estimate could be $\bar{c}_s=\sqrt{{\rm d}p/{\rm d}\rho}\sim
\sqrt{p/\rho}\sim \sqrt{\rho c^2/\rho}=c$.
$n_{\rm c}=3\epsilon n_0/A$, where $\epsilon$ denotes the baryon number density of solid quark matter in the
unit of $n_0$. $n_0$ is the baryon number density of normal nuclear matter and equals 0.17 fm$^{-3}$. We
consider $\epsilon=3-5$ could be the typical values for a SQS. Note that in the following calculation in this
paper, we will adopt $\epsilon=3$, since the variation of $\epsilon$ between $3-5$ would not cause the results
to vary in orders. $A$ is the number of valence quarks in a quark cluster. We may expect that $A\sim10$, since
$A=18$ if quark-$\alpha$-like clusters are formed~\citep{alpha,xu03}, and $A$ could even be conjectured to be in
the order of $\geq 10^2$.
Debye temperature $\theta_D$ of a SQS is then $\sim 10^{12}$ K, which could be even higher than the temperature
when a quark star is born. Hence, the heat capacity in the low-temperature limit or the temperature-cube law, is
applicable for the lattice in the third stage, i.e.
\begin{equation}
c_v^{\rm l} = \frac{12\pi^4}{5}k_{\rm B}(\frac{T_{\rm s}}{\theta_D})^{3} \label{eq:cvl},
\end{equation}
where $c_v^{\rm l}$ is the heat capacity per classical particle (or quark cluster when referring to the solid
quark matter)\citep{VK89}. Thus $C_v^{\rm l}=N\cdot c_v^{\rm l}$, where $N$ is the total number of clusters in a
star.

For the stellar electron gas, those electrons distributed in the vicinity of the Fermi surface will contribute
significant heat capacity. The electron heat capacity would be evaluated by
\begin{equation}
C_v^{\rm e}\sim N_{\rm e}\cdot\frac{k_{\rm B}T_{\rm s}}{E_F}\cdot k_{\rm B}, \label{eq:cve}
\end{equation}
where $N_{\rm e}$ is the number of electrons in a star, and $E_F$ is the Fermi energy of the degenerate electron
gas. In the extremely relativistic case, $E_F=(\frac{3n_{\rm e}h^3}{8\pi})^{1/3}\cdot c$, in which $n_{\rm e}$
is the number density of electrons.
According to the calculation using the bag model, $n_{\rm e}$ is typically chosen as one part in $10^5$ of the
baryon number density of strange quark matter \citep{zhu04}. In the 3 times nuclear density case, $n_{\rm
e}\sim5\times10^{33}$ cm$^{-3}$. $E_F$ is hence in the order of $\sim10$ MeV.
In Fig. \ref{fig:cv}, the heat capacities of these two components are plotted. As can be seen, the electron heat
capacity is much larger than that of the lattice. Noting that it is really ambiguous about the melting
temperature of a SQS, though we extended the calculation to 10 MeV.

We can now evaluate the cooling time scale for a SQS, if the star is only powered by its residual thermal
energy. We choose the stellar mass of 1.4$M_{\odot}$ as a typical case. We then have
\begin{equation}
-C_v\frac{{\rm d}T_{\rm s}}{{\rm d}t}=4\pi R^2\sigma T_{\rm s}^4+L_{\nu}^{\rm pair}+L_{\nu}^{\rm plasma},
\label{eq:timesl}
\end{equation}
where $R$ is the local stellar radius, $\sigma$ is the Stefan-Boltzmann constant.
On the right hand side of the equation (\ref{eq:timesl}), the first energy loss component origins from the
radiation of thermal photons, the second and the third terms are the neutrino luminosities induced by the pair
production and the interior ultra-relativistic electrons, respectively. The calculation about these two neutrino
luminosities refers to the analytic formulae given by \citet{it89}.
The concerning about the high temperature here makes the consideration on the neutrino cooling necessary.
Although the actual neutrino emissivity of clustered quarks is hitherto unknown, these two components of
neutrino emission could really take place for a SQS. Thus, one could obtain an upper bound on the
residual-heat-powered cooling time scale for a SQS via equation (\ref{eq:timesl}).
The calculation showed that a 1.4$M_{\odot}$-SQS could only be sustained in $\sim$18 d, when it cools down from
$10^{11}$ K to $10^5$ K.
This would be the intrinsic distinction for SQSs from neutron stars. Neutron star cooling is mainly residual
thermal energy powered, while thermal X-ray emission and cooling processes of SQSs could however be sustained by
heating processes.
\subsection{Stellar heating}
Heating processes may play a significant role in the thermal evolution or the visibility of pulsars in the soft
X-ray band. Pulsars with different magnetospheric properties may be undergoing different heating mechanisms.
\subsubsection{Spin origin}
Luminous nonthermal radiation and bright pulsar wind nebulae (PWNe), these observational manifestations may
imply active magnetospheres. Strong star wind or relativistic particle flow would be ejected from the poles of
such a pulsar. A certain amount of backflow of such plasma induce the stellar heating, or an energy input could
take place at the polar caps and disperse to the bulk of the star \citep{CR77,A81,ZH00,WRHZ98}.
The heat flow $H$ could thus be
\begin{equation}
H\sim \frac{L_{\rm SH}-4\pi r_{\rm p}^{2}\sigma T_{\rm p}^{4}}{\pi
r_{\rm p}^{2}}\sim \kappa\frac{T_{\rm p}-T_{\rm s}}{R}, \label{eq:H}
\end{equation}
where $r_{\rm p}\approx R\sqrt{\Omega R/c}$ and $T_{\rm p}$ are the radius and the surface temperature of the
polar caps in the local reference frame respectively. We adopted the one-dimensional approximation for the
gradient of temperature, i.e. $\nabla T\sim (T_{\rm p}-T_{\rm s})/R$.
The luminosity of the return-current stellar heating $L_{\rm SH}$ could be a function of the spin energy loss
rate $\dot{E}$ and could generally be in the form of a power law. Moreover, a phenomenological study on the
bolometric luminosity and $\dot{E}$ reveals that the power index could be 1/2 or 1, namely $L_{\rm
SH}=C\dot{E}^{1/2}$ or $L_{\rm SH}=\eta\dot{E}$ (see Appendix A for the details).

The metal-like interior of a SQS could imply that its thermal conductivity $\kappa$ could be the sum of the
partial components of phonons, electrons, and static impurities, or
\begin{equation}
\kappa=\kappa_{\rm p}+\kappa_{\rm e}+\kappa_{\rm imp},
\end{equation}
where the subscripts `p', `e', and `imp' denote the partial components mentioned above, respectively.
The electron thermal conductivity $\kappa_{\rm e}$ could be the dominant component for the metal-like solid
material \citep{FI81}. So phonon contribution $\kappa_{\rm p}$ would be neglected, and, as a preliminary model,
we omit the partial component of static impurities.
The electron thermal conductivity would be written as
\begin{equation}
\frac{1}{\kappa_{\rm e}}=\frac{1}{\kappa_{\rm ee}}+\frac{1}{\kappa_{\rm pe}},
\end{equation}
where $\kappa_{\rm ee}$ is the partial component contributed by the collision between electrons, while
$\kappa_{\rm pe}$ is that contributed by the collision between phonons and electrons. For their analytic
formulae, we refer to \citet{FI81} with assuming that the results remain hold for SQSs.
\subsubsection{Accretion origin}
Some pulsars, showing steady long-term soft X-ray fluxes, being inert in nonthermal emission, lacking the proves
on the existences of the associated PWNe, (and sometimes) owning excesses of X-ray luminosities relative to the
spin energy loss rates, could alternatively be inactive pulsar candidates. Due to their little magnetospheric
manifestations, SQSs will not suggest the thermal X-ray radiation of these sources is of spin origin.

One possible way to understand the origin of the energy is the accretion in the {\it propeller} regime. In this
regime, a shell of atmosphere of matter may form surrounding the star. Matter closing to the inner boundary of
the shell may interact directly with the rotating stellar magnetosphere, as a result of which most of the matter
will be expelled outward \citep{lip92}. Nevertheless, a certain fraction of the accreting material, described by
the accretion efficiency $\eta_{\rm acc}$, may diffuse starward and fall onto the surface of the star finally.
In the propeller regime, the stellar magnetosphere radius or Alfv\'en radius $r_{\rm m}$ would be on one hand
larger than its corotation radius $r_{\rm co}$, but is on the other hand somewhat smaller than the light
cylinder radius $r_{\rm L}$, i.e.
\begin{equation}
r_{\rm co}=(\frac{GM}{4\pi^2})^{1/3}P^{2/3}\lesssim r_{\rm
m}=(\frac{R^6B_{\rm p}^2}{\dot{M}\sqrt{2GM}})^{2/7}\lesssim r_{\rm
L}=\frac{cP}{2\pi},
\end{equation}
where $G$ is the gravitational constant, $M$, $R$, and $B_{\rm p}$ are the stellar mass, radius, and the
magnetic strength at the poles, respectively. The accretion rate $\dot{M}$ could be scaled by Eddington
accretion rate $\dot{M}_{\rm Edd}$; so $\mu$ could denote the accretion rate in this unit, i.e.
$\dot{M}=\mu\dot{M}_{\rm Edd}$.
During the accretion of the material which can reach the stellar surface eventually, the energy of gravitation
would be released. Furthermore, when the two-flavor baryonic matter impact upon the surface of a SQS, it will
burn into the three-flavor strange quark matter phase, and the latent heat of $\Delta \varepsilon \sim 10$ MeV
-- $\sim 100$ MeV per baryon could be released in the phase transition \citep{Madsen99}.
The luminosity of stellar heating in this situation could then be
\begin{equation}
L_{\rm SH}=\frac{GM\cdot \eta_{\rm
acc}\cdot\dot{M}}{R}+\Delta\varepsilon\frac{\eta_{\rm
acc}\cdot\dot{M}}{m_{\rm p}}, \label{eq:polarpowerD}
\end{equation}
where $m_{\rm p}$ is the mass of a proton.

\section{Observations versus expectations}
\begin{table}
{\tiny
\begin{center}
\begin{tabular}{llccccccc}
\hline \\
No. & Source & t (kyr) & $T_{\rm s,1}^{\infty}$ (MK) & $R_{1}^{\infty}$ (km) & $T_{s,2}^{\infty}$ (MK) &
$R_{2}^{\infty}$
(km) & $L_{\rm bol}^\infty$ ($10^{33}$ ergs s$^{-1}$) & Refs. \\
\hline \\
1 & PSR B0531+21 & 1 & $\lesssim1.85$ & $\gtrsim15$ & -- & --
& $\sim18.8$ & (1) \\
\ & (Crab) \\
2 & PSR J1811-1925 & 2 & $\le1.74$ & -- & -- & -- & -- & (2) \\
\ & (in G11.2-0.3) \\
3 & PSR J0205+6449 & 0.82-5.4 & $\approx1.7$ &
$\approx2.6$ & -- & -- & $\approx0.44$ & (3) \\
\ & (in 3C 58) \\
4 & PSR J1119-6127 & $\sim1.6$ & $2.4_{-0.2}^{+0.3}$ &
$3.4_{-0.3}^{+1.8}$ & -- & -- & $2.0^{+2.5}_{-0.4}$ & (4) \\
\ & (in G292.2-0.5) \\
5 & RX J0822-4300 & 2-5 & $2.61^{+0.30}_{-0.26}$ & $3.29_{-0.74}^{+1.12}$ & $5.04^{+0.28}_{-0.20}$ &
$0.75^{+0.12}_{-0.15}$ & $\approx6.2$ & (5) \\
\ & (in Pup A) \\
6 & PSR J1357-6429 & $\sim7.3$ & $1.7\pm0.2$ & $2.5\pm0.5$ & -- & --
& $\approx0.37$ & (6) \\
7 & RX J0007.0+7303 & 10-30 & $<0.66$ &
$\approx12$ & -- & -- & $<4.0\times10^{-2}$ & (7) \\
\ & (in CTA 1) \\
8 & PSR B0833-45 & 11-25 & $1.06\pm0.03$ & $5.1^{+0.4}_{-0.3}$ &
$2.16^{+0.06}_{-0.07}$ & $0.73^{+0.09}_{-0.07}$ & $0.31^{+0.05}_{-0.04}$ & (8) \\
\ & (Vela) \\
9 & PSR B1706-44 & $\sim17$ & $2.01_{-0.20}^{+0.18}$
& $1.81_{-0.29}^{+0.43}$ & -- & -- & $\approx0.38$ & (9) \\
\ & (in G343.1-02.3) \\
10 & PSR B1823-13 & $\sim21$ & $1.61_{-0.07}^{+0.10}$ & $\sim2.5$ & -- & -- & $\approx$0.30 & (10) \\
11 & PSR J0538+2817 & $30\pm4$ & $2.12_{-0.03}^{+0.04}$
&$1.68\pm0.05$ & -- & -- & $\approx0.46$ & (11) \\
\ & (in S147) \\
12 & PSR B2334+61 & $\sim41$ & $1.62\pm0.23$ &
$1.66_{-0.39}^{+0.59}$ & -- & -- & $\approx0.11$ & (12) \\
\ & (in G114.3+0.3) \\
13 & PSR B1916+14 & $\sim88$ & $1.5^{+1.1}_{-0.6}$ & $0.8^{-0.6}_{+5.2}$ & -- & -- & $\approx0.03$ & (2) \\
14 & PSR B0656+14 & $\sim110$ & $0.91\pm0.05$ & $\approx14$ &
$1.9\pm0.4$ & $\approx0.8$ & $\approx0.96$ & (13) \\
\ & (in Monogem Ring) \\
15 & PSR J0633+1746 & $\sim340$ & $0.5\pm0.01$ &
$8.6\pm1.0$ & $1.9\pm0.3$ & $0.04\pm0.01$ & $\approx3.2\times10^{-2}$ & (14) \\
\ & (Geminga) \\
15$'$ & & & 0.482 & $\approx10$ & -- & -- & & (15) \\
16 & PSR B1055-52 & $\sim540$ & $0.79\pm0.03$ & $12.3_{-0.7}^{+1.5}$
& $1.79\pm0.06$ & $0.46\pm0.06$ & $\approx0.46$ & (14) \\
17 & PSR J2043+2740 & $\sim1200$ & $\approx0.9$ & $\approx2$ & -- &
-- & $\approx2.0\times10^{-2}$ & (16) \\
18 & 1E 1207.4-5209 & $\sim7^{+13}_{-4}$ & $1.90\pm0.01$ & $4.5\pm0.1$ & $3.70\pm0.02$ & $0.83\pm0.02$ &
$\approx2.1$ & (17,18)
\\
\ & (in PKS 1209-51/52) \\
19 & CXOU J232327.9 & 0.3 & $6.14\pm0.46$ & $0.41^{+0.08}_{-0.07}$ &
-- & -- & $1.7^{+1.6}_{-0.9}$ & (19) \\
\ & +584842 (in Cas A) \\
20 & CXOU J085201.4 & $<1.1$ & $4.68\pm0.06$ & $0.28\pm0.01$ & -- &
-- & $0.25\pm0.02$ & (20) \\
\ & -461753 (in G266.2-1.2) \\
21 & PSR J1852+0040 & -- & $5.10\pm3.48$ & $0.9\pm0.2$ & -- & -- &
$3.7\pm0.9$ & (21) \\
\ & (in Kes 79) \\
22 & PSR J1713-3949 & -- & 4.4 & 2.4 & -- & -- & 15 & (22) \\
\ & (in G347.3-0.5) \\
23 & RX J1856.5-3754 & $\sim500$ & $\approx0.74$ & $5.0$ & -- & -- &
$\approx5.2\times10^{-2}$ & (23) \\
24 & RX J0720.4-3125 & $\sim1300$ & $\approx0.94$ & $\approx6.1$ &
-- & -- & $\approx0.21$ & (24) \\
25 & RBS 1223$^a$ & $\sim400$ & 1.04 & 0.8 &
-- & -- & $5.1\times10^{-3}$ & (25) \\
26 & RX J0420.0-5022 & $\sim110$ & $0.66^{+0.29}_{-0.54}$ & 1.4 & --
& -- & 2.7$\times10^{-3}$ & (26) \\
27 & RX J0806.4-4123 & -- & $1.09\pm0.01$ & $\approx$0.6 & -- & -- &
$\approx3.6\times10^{-3}$ & (27) \\
28 & RX J1605.3+3249 & -- & 1.07 & $\approx$1.1 & -- & -- &
$1.1\times10^{-2}$ & (25) \\
29 & RBS 1774$^b$ & -- & 1.04 & $\approx$1.1 & -- & -- &
$1.1\times10^{-2}$ & (25) \\
\hline\\
\end{tabular}
\end{center}
\begin{flushleft}
$^a$ 1RXS J130848.6+212708 \newline
$^b$ 1RXS J214303.7+065419
\end{flushleft}}
\caption{\tiny Observational parameters on X-ray pulsars, including the surface temperature components $T_{\rm
s,1/2}^{\infty}$, the emission size components $R_{1/2}^{\infty}$ and the bolometric X-ray luminosity $L_{\rm
bol}^{\infty}$. These are the values detected nearby the earth and fitted or analyzed by the black body model.
Noting that (i) the age of 1E 1207.4-5209 (source No. 18) is adopted according to the estimate to the associated
SNR, which is given by the reference (17); (ii) the ages of RBS 1223 and RX J0420.0-5022 are assessed by
spindown ages in this work; (iii) the age limits of sources No. 1, 3-9, 11-12, 14-17, 23-24 are from
\citet{Yak08}, while the rest ages are from the corresponding literatures listed in the last column; (iv) for 5
XDINs (No. 25 to 29), the black body parameters are obtained with an assumed distance of 100 pc, as a result of
the lack or uncertain of their distances; (v) the spectral parameters for these sources are from the references
listed in the last column: (1) \citet{WOPEBTS04} (2) \citet{zhu09} (3) \citet{SHVM04} (4) \citet{GKCGP05} (5)
\citet{HB06} (6) \citet{Zav07a} (7) \citet{Halpern04} (8) \citet{Manzali07} (9) \citet{McGowan04} (10)
\citet{pav08} (11) \citet{McGowan03} (12) \citet{McGowan06} (13) \citet{Possenti96} (14) \citet{Deluca05} (15)
\citet{Jackson05} (16) \citet{ZP04} (17) \citet{PZST02} (18) \citet{Deluca04} (19) \citet{Cha01} (20)
\citet{Kar02} (21) \citet{Gotthelf05} (22) \citet{pav04} (23) \citet{Ho07} (24) \citet{Kaplan03} (25)
\citet{Haberl04} (26) \citet{Haberl99} (27) \citet{Haberl02}.}\label{tab:data}
\end{table}
\begin{figure}
\begin{center}
\includegraphics[width=0.7\textwidth]{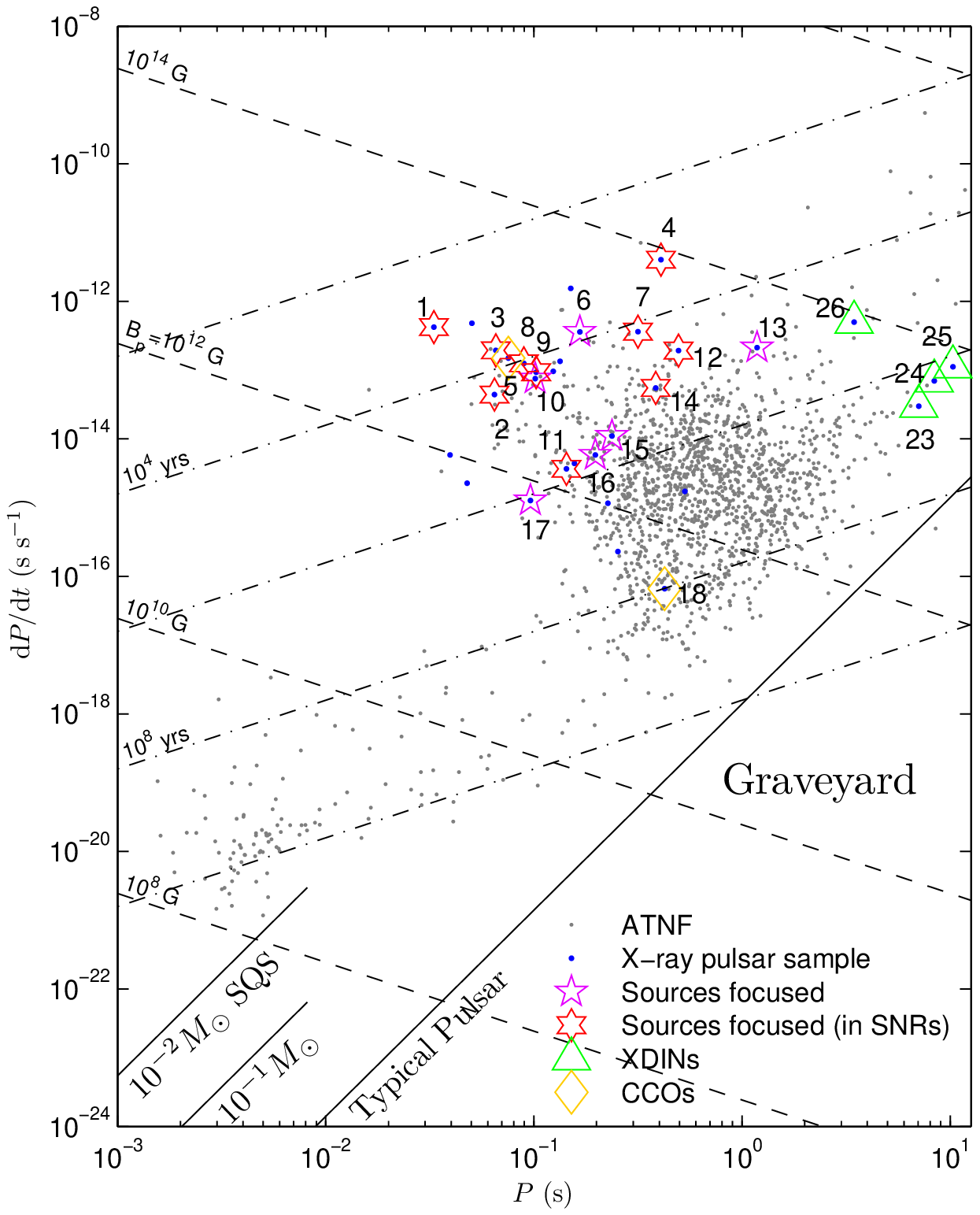}
\caption{$P-\dot{P}$ diagram for the X-ray pulsar sample. The solid lines indicate the death lines both for a
typical pulsar (with mass of 1.4${M_\odot}$ and a radius of 10 km) and low-mass SQSs (with mass of
$10^{-2}-10^{-1}{M_\odot}$), with surface magnetic field $B=10^{12}$ G and potential drop in open field line
region $\delta\phi=10^{12}$ V. The death lines move up if decreasing $B$ and/or increasing
$\delta\phi$~\citep{xu05}. The hexagonals indicate the focused sources associated with detectable supernova
remnants (SNRs), while the pentagons indicate those which exhibit without evident SNRs. The triangles indicate 4
XDINs with measurable temporal parameters, and the diamonds mark the CCOs. The timing parameters for RX
J0822-4300 are from \citet{ZTP99}, while the rest are from ATNF Pulsar Catalog$^1$
.}\label{fig:ppdot}
\end{center}
\end{figure}

We concentrate on those X-ray sources, which demonstrate significant thermal emission, own ordinary magnetic
fields $10^{11-13}$ G, own comparatively young ages $10^{3-6}$ yrs, and have spins of a few tens of milliseconds
to a few seconds.
We mainly refer to the collation made by \citet{Yak08} on cooling neutron stars, by \citet{BA02} on X-ray
pulsars, by \citet{Haberl04} on X-ray dim isolated neutron stars (XDINs), and by \citet{pav04} on central
compact objects (CCOs).
The sample exhibited in Table \ref{tab:data} thus comprises top 17 active pulsar candidates, 7 XDINs (No. 23-29)
and 6 CCOs (No. 5 and 18-22).
We note that the values of the surface temperatures adopted are determined by black body fits, according to the
way that a SQS emits thermal photons.
The XDINs and CCOs are considered to be magnetosphere-inactive pulsar candidates, since their quiescent
manifestations on nonthermal radiation. CCOs, moreover, would only be seen in the soft X-ray band without the
evidences on the existences of the associated PWNe.
Additionally, we note that the emission of the energetic Crab pulsar is much likely to be overwhelmed by the
nonthermal component originating from its luminous plerion, which hampers the detection to the stellar surface
thermal radiation. As an estimate, we, however, adopted a $2\sigma$ upper limit to Crab's surface temperature
and an inferred radius with assuming a 2 kpc distance between Crab and the earth \citep{WOPEBTS04}.
RX J0822-4300, the central stellar remnant in Puppis A, used to be analyzed by \citet{ZTP99} basing on the {\it
ROSAT} observations in 1990s. They could not confirm whether the X-ray structures surrounding the star belong to
the supernova remnant (SNR) or are induced by the probable active magnetosphere. Recent observations on it by
{\it Chandra} and {\it XMM-Newton} telescopes, however, did not reveal the presence of the associated plerion,
indicating an inert magnetosphere \citep{HB06}. Because of the ambiguity of this source, we temporarily make it
as one of the active source candidates.

Furthermore, we emphasize these X-ray sources that own measurable spin parameters in the $P-\dot{P}$ diagram
(Fig. \ref{fig:ppdot}), by which a distribution of them can be read. Other 10 X-ray sources with constraints on
the upper limits of their bolometric luminosities are also denoted in the diagram \citep{BA02}. Besides the
death line for a typical pulsar, those for low-mass SQSs are also indicated in the diagram. These death lines
set boundaries of the `graveyards' for pulsars with different mass~\citep{xu05}. The timing parameters are from
ATNF Pulsar Catalog\footnote{www.atnf.csiro.au/research/pulsar/psrcat.} \citep{Manchester05}, except for RX
J0822-4300, which refers to \citet{ZTP99}.

For magnetosphere-active pulsar candidates, SQSs would suggest themselves to reproduce the cooling processes,
while, for magnetosphere-inactive pulsar candidates, possible approaches to understand their current X-ray
luminosities would also be proposed by SQSs. These two will be presented in \S 3.1 and \S 3.2.
\subsection{Cooling of active pulsar candidates}
\begin{sidewaysfigure}
\begin{center}
\includegraphics[width=1.0\textwidth]{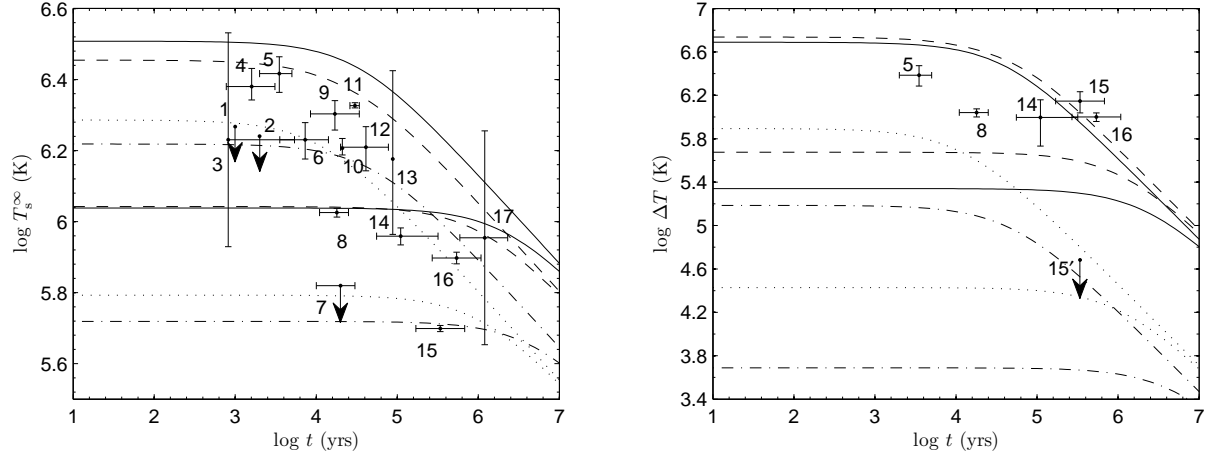}
\caption{{\it Left} panel: Cooling curves for SQSs, if the 1/2-law between the bolometric luminosity and the
spin energy loss rate holds. {\it Right} panel: Corresponding temperature differences between the hot and warm
components of SQSs, or $\Delta T=T^{\infty}_{\rm p}-T^{\infty}_{\rm s}$. The parameters in both panels:
$M=0.1{M_\odot}$, $C=10^{16}$ (solid lines); $M=1.0{M_\odot}$, $C=10^{16}$ (dashed lines); $M=1.0{M_\odot}$,
$C=10^{15}$ (dotted lines); $M=0.01{M_\odot}$, $C=10^{15}$ (dash-dot lines) ($C$ is in the unit of ergs$^{1/2}$
s$^{-1/2}$). For two curves with same $M$ and $C$, the upper one corresponds to an initial spin of 10 ms, while
that of the lower one is 100 ms. Noting that the errors on the surface temperatures of PSRs J0205+6449 (No. 3)
and J2043+2740 (No. 17) are not provided by the authors of the references, we then conservatively adopt them as
deviating from the central values by a factor of 2.}\label{fig:ccos}
\end{center}
\end{sidewaysfigure}
\begin{sidewaysfigure}
\begin{center}
\includegraphics[width=1.0\textwidth]{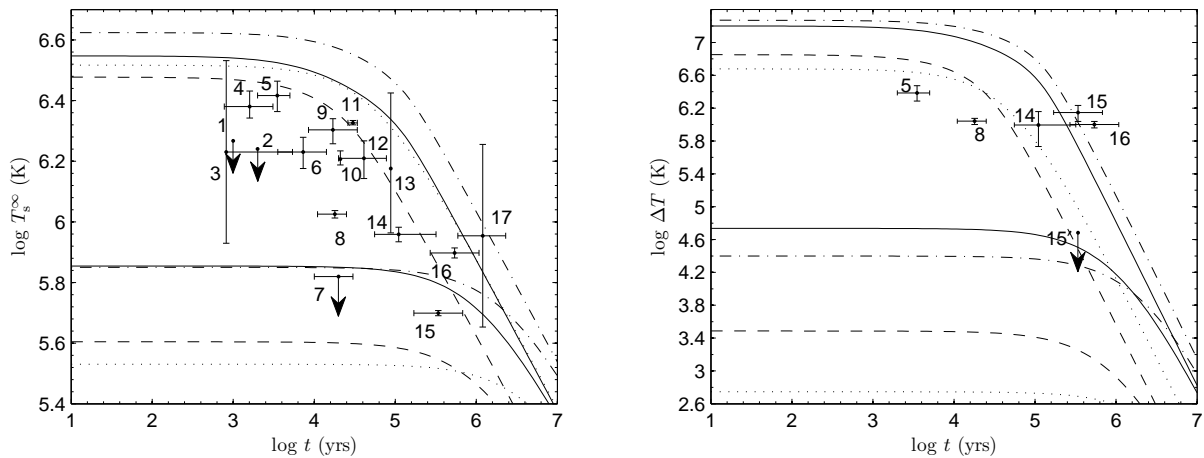}
\caption{{\it Left} panel: Cooling curves for SQSs, if the linear-law holds. {\it Right} panel: Temperature
differences for this case. The parameters: $M=1.0{M_\odot}$, $\eta=0.01$ (solid lines); $M=1.0{M_\odot}$,
$\eta=0.001$ (dashed lines); $M=0.1{M_\odot}$, $\eta=0.1$ (dash-dot lines); $M=0.01{M_\odot}$, $\eta=0.1$
(dotted lines). As in Fig. \ref{fig:ccos}, for two curves with same $M$ and $\eta$, the upper one corresponds to
an initial spin of 10 ms, while the lower one 100 ms.}\label{fig:cclr}
\end{center}
\end{sidewaysfigure}
\begin{table}
{\tiny
\begin{center}
\begin{tabular}{clllcc}
\hline \\
No. & Source & $\nu$ (s$^{-1}$) & $\dot{\nu}$ (s$^{-2}$) &
$I_{45}^{a}\cdot C^{2}$ (ergs s$^{-1}$) & $I_{45}\cdot\eta$ \\
\hline \\
1 & PSR B0531+21 (Crab) & 30.225437 & -3.862$\times10^{-10}$ &
7.7$\times10^{29}$ & 4.1$\times10^{-5}$ \\
2 & PSR J1811-1925 & 15.463838 & -1.052$\times10^{-11}$ & -- & -- \\
3 & PSR J0205+6449 & 15.223856 & -4.495$\times10^{-11}$ &
5.8$\times10^{28}$ & 4.6$\times10^{-5}$ \\
4 & PSR J1119-6127 & 2.452508 & -2.419$\times10^{-11}$ &
1.7$\times10^{30}$ & 8.5$\times10^{-4}$ \\
5 & RX J0822-4300  & 13.2856716499(3) &
-2.6317(3)$\times10^{-11}$ & 2.7$\times10^{30}$ & 2.8$\times10^{-4}$ \\
6 & PSR J1357-6429 & 6.020168 & -1.305$\times10^{-11}$ &
4.4$\times10^{28}$ & 1.2$\times10^{-4}$ \\
7 & RX J0007.0+7303 & 3.165922 & -3.623$\times10^{-12}$
& 3.5$\times10^{27}$ & 8.8$\times10^{-5}$ \\
8 & PSR B0833-45 (Vela) & 11.194650 & -1.567$\times10^{-11}$ &
1.4$\times10^{28}$ & 4.5$\times10^{-5}$ \\
9 & PSR B1706-44 & 9.759978 &
-8.857$\times10^{-12}$ & 4.2$\times10^{28}$ & 1.1$\times10^{-4}$ \\
10 & PSR B1823-13 & 9.855532 & -7.291$\times10^{-12}$ & 3.2$\times10^{28}$ & 1.1$\times10^{-4}$ \\
11 & PSR J0538+2817 & 6.985276 & -1.790$\times10^{-13}$ &
4.3$\times10^{30}$ & 9.3$\times10^{-3}$ \\
12 & PSR B2334+61 & 2.018977 &
-7.816$\times10^{-13}$ & 1.8$\times10^{29}$ & 1.7$\times10^{-3}$ \\
13 & PSR B1916+14 & 0.846723 & -1.523$\times10^{-13}$ & 1.8$\times10^{29}$ & 5.9$\times10^{-3}$ \\
14 & PSR B0656+14 & 2.598137 & -3.713$\times10^{-13}$ &
1.1$\times10^{31}$ & 1.7$\times10^{-2}$ \\
15 & PSR J0633+1746 (Geminga) & 4.217640 & -1.952$\times10^{-13}$ &
3.2$\times10^{28}$ & 9.9$\times10^{-4}$ \\
16 & PSR B1055-52 & 5.073371 & -1.501$\times10^{-13}$ &
6.9$\times10^{30}$ & 1.5$\times10^{-2}$ \\
17 & PSR J2043+2740 & 10.402519 & -1.374$\times10^{-13}$ &
7.1$\times10^{27}$ & 3.5$\times10^{-4}$ \\
\hline \\
\end{tabular}
\end{center}
\begin{flushleft}
$^a$ $I_{45}=I/(10^{45}$ g cm$^2$)
\end{flushleft}}
\caption{The moments of inertia of active pulsar candidates suggested by SQSs. The 5th column lists the values
derived by the 1/2-law, while the 6th column lists those obtained by the linear-law. The temporal parameters are
from ATNF Pulsar Catalog (see footnote 1 for the website). For RX J0822-4300, we note that we adopt the spin
parameters obtained by \citet{ZTP99} basing on a 4.5 yr-span data set, though \citet{HB06} provided more recent
values using a 0.5 yr-span data set.}\label{tab:moi}
\end{table}
If two isotropic black body emission components --- the hot component for polar caps and the warm one for the
bulk of the star --- are defined for SQSs, then the equation describing cooling processes would be written as
(cf. equation (\ref{eq:coolinggeneral}))
\begin{equation}
L_{\rm SH}=4\pi R^2\sigma T_{\rm s}^{4}+4\pi r_{\rm p}^2\sigma
T_{\rm p}^4, \label{eq:coolingactive}
\end{equation}
and the relation between $T_{\rm s}$ and $T_{\rm p}$ is given by equation (\ref{eq:H}). The luminosity of
stellar heating $L_{\rm SH}$ could either follow the 1/2-law or the linear-law (see Appendix A). Cooling
behaviors would thus be calculated by assuming SQSs rotate as orthogonal rotators and slow down as a result of
magnetic dipole radiation with magnetic field strength at the poles of $10^{12}$ G. We provided a parameter
space to fit the observational temperature-age data, and the comparison between the observations and
expectations are shown in Figs. \ref{fig:ccos} (for 1/2-law case) and \ref{fig:cclr} (for linear-law case).
Some pulsars demonstrate two black body components in their thermal spectra, e.g. Vela pulsar and the Three
Musketeers, implying that the temperature inhomogeneity on these pulsars could be significant. We thus carried
out a temperature-difference fit at the same time, as have been shown in the {\it right} panels in both Figs.
\ref{fig:ccos} and \ref{fig:cclr}. It is worthy of noting that, for Geminga, if the photon index of the
power-law (PL) component is thawed when fitting its phase-resolved spectra, the entry of the hot black body
component could not improve the fits remarkably, or it may even becomes an artifact. This may mean the
fluctuation of the magnetospheric emission during a spin cycle might mislead the understanding on its X-ray
spectra, as analyzed by \citet{Jackson05} and their results are denoted by No. 15$'$ in Table \ref{tab:data}. If
this situation holds, the surface temperature fluctuation on Geminga could be tiny so that undetectable. We,
hence, set a rough upper limit to the temperature difference for Geminga by an order lower than its surface
temperature, as denoted by 15$'$ in the `temperature difference-age' figures.

The proposed relation between the rotational kinetic energy loss rate and the bolometric luminosity could
inversely provide direct measurement to the moments of inertia of the active X-ray sources. The results are
exhibited in Table \ref{tab:moi} and could be understood in the SQS regime, especially low-mass SQSs (cf. the
adoption of the coefficients $C$ and $\eta$ in Figs. \ref{fig:ccos} and \ref{fig:cclr}). We note that for SQSs
with 0.01${M_\odot}$, 0.1${M_\odot}$, 1.0${M_\odot}$ and 2.0${M_\odot}$, the moments of inertia $I_{45}$ (i.e.
values scaled by $10^{45}$ g cm$^{2}$) are $\sim2.55\times10^{-4}$, $\sim$0.01, $\sim$0.55 and $\sim$1.74
respectively.
\subsection{Thermal X-ray emission of inactive pulsar candidates}
\begin{table}
\begin{center}
{\scriptsize
\begin{tabular}{ccc}
\hline \\
Mass & $\frac{\dot{M}}{\mu} (10^{-9}$ ${M_\odot}$ yr$^{-1}$) &
$\frac{L_{\rm bol}}{\mu\cdot \eta_{\rm acc}} (10^{38}$ ergs s$^{-1}$) \\
\hline \\
$10^{-2}{M_\odot}$ & 2.7 & 1.7$\times10^{-1}$ \\
$10^{-1}{M_\odot}$ & 5.7 & 4.7$\times10^{-1}$ \\
$1.0{M_\odot}$ & 12.4 & 2.0 \\
\hline \\
\end{tabular}}
\end{center}
\caption{The bolometric X-ray luminosity ($L_{\rm bol}$) contributed by the accretion of a SQS in the {\it
propeller} regime. The corresponding accretion rates $\dot{M}=\mu\dot{M}_{\rm Edd}$ are also given. Noting that
the values of luminosities are calculated by adopting $\Delta\varepsilon$, the latent heat per baryon during the
phase transition, equals 100 MeV. In the propeller phase, the fraction of $\eta_{\rm acc}$ of the accreted
matter diffuses starward and falls onto the stellar surface eventually.}\label{tab:lum}
\end{table}
As has been discussed in \S 2.3.2, one probable energy source to be responsible for the soft X-ray emission of
magnetosphere-inactive pulsar candidates lies in the accretion in the {\it propeller} regime. The accretion
could be either to the interstellar medium or to the fallback disk. In this regime, however, the stellar
magnetosphere radius could hardly be given, even though the sources' spindown rates are detectable. Because the
spindown rate in this case would be dominated by the accretion behavior.
Equation (\ref{eq:coolinggeneral}) for this case could be specified as
\begin{equation}
L_{\rm bol}=\frac{GM\cdot \eta_{\rm
acc}\cdot\dot{M}}{R}+\Delta\varepsilon\frac{\eta_{\rm
acc}\cdot\dot{M}}{m_{\rm p}}, \label{eq:coolingdead}
\end{equation}
with considering equation (\ref{eq:polarpowerD}) and the stellar residual thermal energy is too small to be
taken into account.
Table \ref{tab:lum} lists the luminosities of SQSs under the accretion scenario. It could be concluded that the
sources' current observational thermal X-ray luminosities would be interpreted by considering the parameters
$\mu$, the accretion rate in the unit of that of Eddington, and $\eta_{\rm acc}$, the accretion efficiency, such
as $\mu=0.01$, $\eta_{\rm acc}=0.001$.
\section{Conclusions and discussions}
We collate the thermal observations of 29 X-ray isolated pulsars and, in the SQS regime, for the
magnetospherically active pulsar candidates, establish their cooling processes (Figs. \ref{fig:ccos} and
\ref{fig:cclr}), while for the magnetospherically inactive or dead pulsar candidates, interpret the X-ray
luminosities under the accretion scenario. (Table \ref{tab:lum}).
Fitting to the thermal spectra for X-ray pulsars using the black body model often results in such small emission
sizes that could naturally be interpreted by the SQS model. A recent study on the Cassiopeia A CCO shows that
the fitted emission size is still significantly smaller than a typical neutron star radius, even if the hydrogen
atmosphere model is employed in the fitting \citep{luna09}.
The uncertain estimates on the distances of pulsars may introduce considerable errors in the fitting about the
X-ray source emission sizes.
However, if these obtained emission radii can be trusted, SQSs, because of the possibility of being low-mass,
could provide an approach to understand these observational manifestations. We note that for SQSs with mass of
$0.01{M_\odot}$, $0.1{M_\odot}$ and $1{M_\odot}$, their radii are $\sim$1.8, $\sim$3.8 and $\sim$8.3 km
respectively.
On the other hand, a linkage between pulsar rotational kinetic energy loss rates and bolometric X-ray
luminosities is explored by SQSs (see Appendix A), and the resulting estimates on pulsar moments of inertia is
exhibited in Table \ref{tab:moi}.

We hence conclude that the phenomenological SQS pulsar model could not be ruled out by the thermal observations
on X-ray isolated pulsars, though a full depiction on the
thermal evolution for a quark star in all stages could hardly be given nowadays, as the lack of the physics in
some extreme conditions (as has been discussed in \S 2.1).
SQSs have significant distinguishable interiors with neutron stars, but the structures of the magnetosphere
between these two pulsar models could be similar. Therefore, SQSs and neutron stars would have similar heating
mechanisms, such as the bombardment by backflowing particles and the accretion to surrounding medium. However, a
full comparison between SQSs and neutron stars including their cooling processes as well as the heating mechanisms
has beyond the scope of this paper, and should be interesting and necessary in the future study.

The various performances of X-ray pulsars may indicate their current states or properties and imply their
possible evolutionary history.
In addition, if these X-ray sources are actually SQSs, the possible formation mechanism could be an interesting
topic.
We thus extend a discussion in \S 4.3.
\subsection{Spin-powered pulsars}
Spin has always been a significant energy source for active pulsars, by which multiwave bands nonthermal
radiation are driven, including those of the pulsar themselves as well as those of the surrounding plerions. Some
phenomenological studies demonstrate certain regularities for such an active pulsar population. A brief summary is that the
nonthermal X-ray luminosity and spin energy loss rate own a relation of $L_{\rm X}=10^{-3}\dot{E}$ \citep{BT97},
and the $\gamma$-ray luminosity is proportional to the square root of $\dot{E}$ or
$L_{\gamma}\propto\dot{E}^{1/2}$ \citep{Thomp97}. We, additionally, note that $V\propto\dot{E}^{1/2}$, where $V$
is the potential drop along the open field region. Therefore, it seems that for such a population the more
younger the larger $\dot{E}$ and $V$ that a pulsar owns (cf. Fig. \ref{fig:edot}, {\it bottom} panel) and the
more luminous of the radiation in the hard X-ray and $\gamma$-ray bands.

Besides nonthermal emission, spin may also be an energy origin for pulsar thermal radiation as the relations of
$L_{\rm bol}^{\infty}\propto\dot{E}^{1/2}$ or $L_{\rm bol}^{\infty}\sim10^{-3}\dot{E}$ could exist
observationally (see Appendix A for details).
SQSs predominantly follow these relations to accomplish their cooling processes, since their residual thermal
energy could be quite inadequate to sustain a long term X-ray thermal emission.
\subsection{Accretion-powered pulsars}
XDINs and CCOs could be representative populations of magnetosphere-inactive pulsars. Nevertheless, their energy
origin that could power the X-ray emission is still an enigma. The bolometric luminosities of 1E 1207.4-5209, RX
J1856.5-3754 and RX J0720.4-3125 exceed their spin energy loss rates by a factor of $\sim60$, $\sim15$ and
$\sim45$, respectively; they appear not to be powered by spins.

CCOs could be a group of weakly-magnetized and long-initial-period pulsars \citep{PZST02,Gotthelf07}, because of
which their potential drops along the open field lines could be much less than $\sim10^{12}$ V, so that the
primary plasma could not be fully accelerated, resulting in inactive magnetospheres.
The black body fits to the thermal spectra of CCOs always result in extremely small emission sizes
\citep{pav04}, implying the possible existence of low-mass SQSs. Therefore, the $\dot{E}$ of CCOs might be very
low, so that effective magnetospheric emission and plerions could not be driven.
CCOs could thus be the representatives of a group of natal inactive pulsars.
Accretion to the fallback ejecta of the associated SNRs could power their soft X-ray luminosities, and this
scenario have been mentioned by several authors \citep[e.g.,][]{Cha01,Kar02,Fe06,Halpern07}.

XDINs, owning comparatively larger spindown ages and stronger magnetic fields, might be the descendants of
magnetars. Considering similar properties between them \citep{Haberl04,Mere08}, one could not exclude that XDINs
are still being powered by the decaying magnetic fields.
However, observations have revealed the probable existence of residual disks around such pulsars, and thus their
radiation may be of accretion origin \citep{Lo07}.
In this case, XDINs could be the evolved products of active pulsars, and thanks to the accretion so that they
are still visible after their death.
Most XDINs have absorption features in their spectra, being similar to those of CCOs. Hence, if they are really
undergoing accretion in the propeller regime, their spindown rates $\dot{P}$ could then be of accretion origin
rather than causing by the magnetic dipole braking. Thus their magnetic fields could be much lower than the
values determined by canonical magnetic dipole radiation, since the absorption features could be electron
cyclotron lines. Therefore, the common properties of absorption features between XDINs and CCOs could imply that
XDINs might be older CCOs.
XDINs and CCOs have not manifested themselves as radio pulsars, their radio-quiet demonstrations could be an
intrinsic property rather than being of beaming origin, since they might be dead or natal inactive pulsars. So
they could be grouped as radio-quiescent pulsars. For RX J0007.0+7303, an active pulsar candidate, its lack of
radio signal could be the result of an unfavorable geometry, or its radio beam sweeps away from the earth as has
been analyzed by \citet{BJ99}.

The similar thermal manifestations of accreting pulsars (e.g. XDINs and CCOs) and cooling pulsars may have
caused a confusion about distinguishing between the two classes \citep{tre00}. However, the activeness of the
magnetosphere could provide a way to achieve an identification. If pulsars are in fact SQSs with rapid rotation,
cooling pulsars are then likely to be undergoing the spindown-powered heating evolution, and multi-bands
nonthermal emission originating from luminous magnetospheres and even PWNe will accompany such coolers during
the processes. In contrast, accreting pulsars may have not demonstrated such characteristics observationally. In
this context, the CCO in the SNR Cassiopeia A (source No. 19 in Table \ref{tab:data}) should be an accretor
because of the lack of PWN and nonthermal power-law component \citep{luna09}.
\subsection{Formation of low-mass quark stars}

Possible low-mass compact stars is a direct consequence of the suggestion that pulsars could be quark
stars~\citep{xu05}. However, the probable existence of quark stars, spanning large mass range, could naturally
raise such a question: how did these stars be created? Indeed, this could be another severe problem given by the
observations.
Being different from solar-mass quark stars (which could originate from core-collapse supernovae), low-mass
quark stars could be the central remnants left by the detonation of the accretion-induced collapse (AIC) of
white dwarfs (WDs) \citep{xu10}. The combustion from the hadronic matter to the strange quark matter could be
the engine of such an explosion. During this phase transition, as high as $\sim100$ MeV per baryon could be
released as the latent heat \citep{Madsen99}. The combustion could hence be in such a high efficiency that \%10
of the rest mass is liberated.
For a WD approaching the Chandrasekhar limit, its mass and radius are $M_{\rm wd}\sim1.4M_{\odot}$, $R_{\rm
wd}\sim10^8$ cm respectively. The gravitational energy is then $E_{\rm g}\sim(3/5)GM_{\rm wd}^2/R_{\rm
wd}\simeq3\times10^{51}$ ergs.
For a successful explosion of such a WD, the necessary mass of a quark star $M_{\rm qs,min}$ would hence be
obtained by
\begin{equation}
0.1M_{\rm qs,min}c^2\simeq E_{\rm g}, \label{eq:Mqsmin}
\end{equation}
as $M_{\rm qs,min}\simeq 2\times10^{-2}M_{\odot}$.
So, in this case, low-mass quark stars with a few tenth of or a few hundredth of solar mass could form, although
one should be aware that the exact value of the quark star mass depends on the region where the detonation wave
inside the WD can sweep though as well as the latent heat of the phase transition, which would be strongly
model-dependent.
These nascent quark stars would keep bare if they has not suffered massive Super-Eddington accretion
\citep{xu02}. An accreted ion (e.g. a proton) should gain enough kinetic energy to penetrate the Coulomb barrier
of a quark star with mass $M_{\rm qs}$ and radius $R_{\rm qs}$, or $GM_{\rm qs}m_p/R_{\rm qs}>\mathcal{V}_{\rm
q}$. The Coulomb barrier $\mathcal{V}_{\rm q}$ is model-dependent, which could varies from $\sim20$ MeV to even
$\sim0.2$ MeV \citep{xu10,xq99}. Considering $M_{\rm qs}=(4/3)\pi R_{\rm qs}^3\rho$ for low-mass quark stars, a
mass lower-limit $M_{\rm qs,lim}$ would then be determined as
\begin{equation}
M_{\rm qs}>M_{\rm qs,lim}=\sqrt{\frac{3\mathcal{V}_{\rm q}^3}{4\pi
G^3m_p^3\rho}}\simeq4.6\times10^{-4}\mathcal{V}_{q,1}^{3/2}\rho_3^{-1/2}M_{\odot}, \label{eq:Mqslim}
\end{equation}
where the density $\rho=\rho_3\times(3\rho_0)$, with $\rho_0$ the nuclear density, and $\mathcal{V}_{\rm
q}=\mathcal{V}_{\rm q,1}$ MeV.
Apart from the formation of the low-mass quark stars discussed here, other astrophysical processes, such as the
collision between two strange stars, could also be the potential sources to give birth to the quark stars with
smaller mass than that of the sun \citep{alc86}.

\appendix
\section{Spin-powered thermal emission?}
\begin{figure}
\begin{center}
\includegraphics[width=0.7\textwidth]{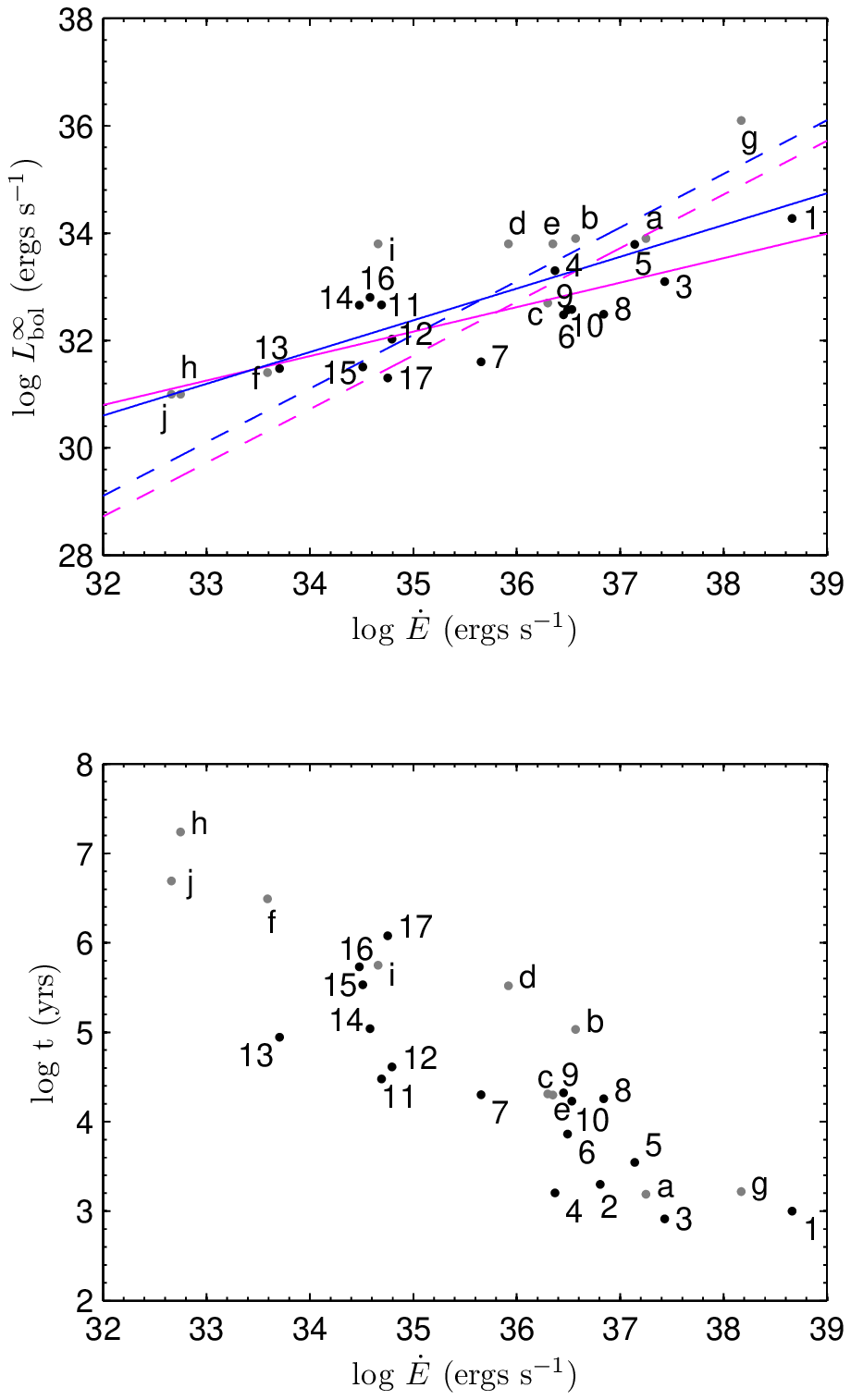}
\caption{Functions $L_{\rm bol}^{\infty}(\dot{E})$ ({\it top} panel) and $t(\dot{E})$ ({\it bottom} panel) of
active pulsar candidates. In the {\it top} panel, the fits are carried out both for Group A (including top 17
pulsars listed in Table \ref{tab:data}; they are marked by dark points and their numbers) and Group B (including
the members in Group A and the other 10 sources whose upper limits on the bolometric luminosities could be
defined observationally; the 10 sources are marked by grey points and letters). Red lines are the fitted results
for Group A, while blue lines are those for Group B. In both groups, solid lines provide the best fits to the
data, while the dashed lines give the fits by freezing $p1$ at 1. We note that the 10 sources with upper limits
on their bolometric luminosities defined are taken from \citet{BA02}, and they are a. B1509-58, b. B1951+32, c.
B1046-58, d. B1259-63, e. B1800-21, f. B1929+10, g. B0540-69, h. B0950+08, i. B0355+54, j.
B0823+26.}\label{fig:edot}
\end{center}
\end{figure}
\begin{table}
{\scriptsize
\begin{center}
\begin{tabular}{ccccc}
\hline \\
Group & $p1$$^a$ & $p2$ & Corr. Coef.$^b$ & $\chi_r^2$ (d.o.f)$^c$ \\
\hline \\
A. & $0.4561\pm0.2315$ & $16.20\pm8.30$ & -- & 0.3277(14) \\
 & (1) & $-3.280^{+0.495}_{-0.494}$ & 0.7487 & 0.8607(15) \\
B. & $0.5918^{+0.2045}_{-0.2046}$ & $11.66^{+7.3100}_{-7.2990}$ &
-- & 0.6081(24) \\
 & (1) & $-2.896\pm0.403$ & 0.7730 & 0.9964(25) \\
\hline \\
\end{tabular}
\end{center}
\begin{flushleft}
$^a$ Values in the parenthesis are frozen during the fits. The errors are in 95\% confidence level.
\newline
$^b$ Correlation coefficient of the data between bolometric luminosity and $\dot{E}$.\newline
$^c$ Reduced $\chi^2$, or $\chi^2$ per degree of freedom (d.o.f).
\caption{Fitting parameters for $L_{\rm bol}^{\infty}$---$\dot{E}$ data, which has been shown in Fig.
\ref{fig:edot} ({\it top} panel).}\label{tab:fit}
\end{flushleft}}
\end{table}
In the solid quark star(SQS) regime, cooling of magnetosphere-active pulsars depends predominantly on the
stellar heating as a result of the lack of sufficient residual heat as has been analyzed in \S 2.2. The pulsar
activity induced stellar heating would have intrinsically set up a linkage between the X-ray bolometric
luminosity and the spin energy loss rate. We, hence, present here a phenomenological study on such a relation.

The sample of active pulsars firstly includes the top 17 sources listed in Table \ref{tab:data}. \citet{BA02}
summarized X-ray pulsars, in which other 10 sources' upper limits on the bolometric luminosities could be
defined. These 10 pulsars are meanwhile considered.
Fig. \ref{fig:edot} ({\it top} panel) illustrates these sources' bolometric luminosities $L_{\rm bol}^{\infty}$
as a function of their spin energy loss rates $\dot{E}$, and Table \ref{tab:fit} lists the fitted parameters led
by the function of
\begin{equation}
{\rm log}L_{\rm bol}^{\infty}=p1\cdot {\rm log}\dot{E}+p2. \label{eq:fit}
\end{equation}
The fits are carried out both only for the top 17 pulsars taken from Table \ref{tab:data} (as Group A) and for
the whole sample with the other 10 sources included at the same time(as Group B).
The best fits, moderate clearly, suggest a 1/2-law on the relation between pulsar bolometric X-ray luminosities
and the spin energy loss rates, or
\begin{equation}
L_{\rm bol}^{\infty}(\dot{E})=C\dot{E}^{1/2}, \label{eq:onesecond}
\end{equation}
with a coefficient $C=10^{p2}$ in the unit of ergs$^{1/2}$ s$^{-1/2}$.
If $p1$ is fixed at 1, a linear-law is then implied, i.e.
\begin{equation}
L_{\rm bol}^{\infty}(\dot{E})=\eta\cdot\dot{E}, \label{eq:linear}
\end{equation}
with a coefficient $\eta=10^{p2}$, or $\eta$ is the conversion efficiency. The fits obtain $\eta\sim10^{-3}$,
which is similar to the nonthermal X-ray case \citep{BT97}. For the 1/2-law case, the conversion efficiency
turns out to be the function of $\dot{E}$, which would be
\begin{equation}
\eta(\dot{E})=\frac{C\dot{E}^{1/2}}{\dot{E}}=\frac{C}{\dot{E}^{1/2}},
\label{eq:eta}
\end{equation}
which could in turn, besides the fits, provide a reference to the value of the coefficient $C$ by a natural
constraint of $\eta(\dot{E})<1$. Taking B0823+26, the minimum $\dot{E}$ in our sample as an example, its $C$
would be less than $\sim\sqrt{10^{32}}=10^{16}$ ergs$^{1/2}$ s$^{-1/2}$. Additionally, we note here that we also
present a age-$\dot{E}$ relation in the {\it bottom} panel of Fig. \ref{fig:edot}, which illustrates a natural
trend for spin-powered pulsars.
\section*{Acknowledgement}
The authors are grateful to Prof. Fredrick Jenet for his valuable comments and advices, especially for his
efforts on improving the English expression of this paper. We wish to thank Mr. Weiwei Zhu for his detailed
introduction to the observational situation about the sources PSR J1811-1925 and PSR B1916+14. The authors also
thank the colleagues in the pulsar group of Peking University for the helpful discussion. This work is supported
by NSFC (10778611, 10973002), the National Basic Research Program of China (Grant 2009CB824800), and by LCWR
(LHXZ200602).


\begin{thebibliography}{99}
%
\bibitem[\protect\citeauthoryear{Alcock, Farhi \& Olinto}{1986}]{alc86}
Alcock C., Farhi E., Olinto A., 1986, ApJ, 310, 216

\bibitem[\protect\citeauthoryear{Alford et al.}{2008}]{csc08}
Alford M. G., Rajagopal K., Schaefer T., Schmitt A., 2008, Rev. Mod. Phys. in press (arXiv:0709.4635)

\bibitem[\protect\citeauthoryear{Arons}{1981}]{A81}
Arons J. 1981, ApJ, 248, 1099

\bibitem[\protect\citeauthoryear{Becker \& Aschenbach}{2002}]{BA02}
Becker W., Aschenbach B., 2002, in WE-Heraeus Seminar on Neutron
Stars, Pulsars and Supernova Remnants, Eds. W. Becker, H. Lesch \&
J. Tr$\ddot{\rm u}$mper, Bad Honnef, pp.64

\bibitem[\protect\citeauthoryear{Becker \& Tr$\ddot{\rm u}$mper}{1997}]{BT97}
Becker W., Tr$\ddot{\rm u}$mper J., 1997, A\&A, 326, 682

\bibitem[\protect\citeauthoryear{Brazier \& Johnston}{1999}]{BJ99}
Brazier K. T. S., Johnston S., 1999, MNRAS, 305, 671

\bibitem[\protect\citeauthoryear{Chakrabarty et al.}{2001}]{Cha01}
Chakrabarty D., Pivovaroff M. J., Hernquist L. E., Heyl J. S.,
Narayan R., 2001, ApJ, 548, 800

\bibitem[\protect\citeauthoryear{Cheng \& Ruderman}{1977}]{CR77}
Cheng A. F., Ruderman M. A., 1977, ApJ, 214, 598

\bibitem[\protect\citeauthoryear{De Luca et al.}{2004}]{Deluca04}
De Luca A., Mereghetti S., Caraveo P. A., Moroni M., Mignani R. P.,
Bignami G. F., 2004, A\&A, 418, 625

\bibitem[\protect\citeauthoryear{De Luca et al.}{2005}]{Deluca05}
De Luca A., Caraveo P. A., Mereghetti S., Negroni M., Bignami G. F.,
2005, ApJ, 623, 1051

\bibitem[\protect\citeauthoryear{Fesen, Pavlov \& Sanwal}{2006}]{Fe06}
Fesen R. A., Pavlov G. G., Sanwal D., 2006, ApJ, 636, 848

\bibitem[\protect\citeauthoryear{Flowers \& Itoh}{1981}]{FI81}
Flowers E., Itoh N., 1981, ApJ, 250, 750

\bibitem[\protect\citeauthoryear{Gonzalez et al.}{2005}]{GKCGP05}
Gonzalez M. E., Kaspi V. M., Camilo F., Gaensler B. M., Pivovaroff
M. J., 2005, ApJ, 630, 489

\bibitem[\protect\citeauthoryear{Gotthelf \& Halpern}{2007}]{Gotthelf07}
Gotthelf E. V., Halpern J. P., 2007, ApJ, 664, L35

\bibitem[\protect\citeauthoryear{Gotthelf, Halpern \& Seward}{2005}]{Gotthelf05}
Gotthelf E. V., Halpern J. P., Seward F. D., 2005, ApJ, 627, 390

\bibitem[\protect\citeauthoryear{Haberl, Pietsch \& Motch}{1999}]{Haberl99}
Haberl F., Pietsch W., Motch C., 1999, A\&A, 351, L53

\bibitem[\protect\citeauthoryear{Haberl \& Zavlin}{2002}]{Haberl02}
Haberl F., Zavlin V. E., 2002, A\&A, 391, 571

\bibitem[\protect\citeauthoryear{Haberl}{2004}]{Haberl04}
Haberl F., 2004, Advances in Space Research, 33, 638

\bibitem[\protect\citeauthoryear{Halpern et al.}{2004}]{Halpern04}
Halpern J. P., Gotthelf E. V., Camilo F., Helfand D. J., Ransom S.
M., 2004, ApJ, 612, 398

\bibitem[\protect\citeauthoryear{Halpern et al.}{2007}]{Halpern07}
Halpern J. P., Gotthelf E. V., Camilo F., Seward F. D., 2007, ApJ,
665, 1304

\bibitem[\protect\citeauthoryear{Ho et al.}{2007}]{Ho07}
Ho W. C. G., Kaplan D. L., Chang P., Van Adelsberg M., Potekhin A.
Y., 2007, MNRAS, 375, 821

\bibitem[\protect\citeauthoryear{Horvath}{2005}]{Horvath05}
Horvath J., 2005, Mod. Phys. Lett., A20, 2799

\bibitem[\protect\citeauthoryear{Hui \& Becker}{2006}]{HB06}
Hui C. Y., Becker W., 2006, A\&A, 454, 543

\bibitem[\protect\citeauthoryear{Itoh et al.}{1989}]{it89}
Itoh N., Adachi T., Nakagawa M., Kohyama Y., 1989, 339, 354

\bibitem[\protect\citeauthoryear{Jackson \& Halpern}{2005}]{Jackson05}
Jackson M. S., Halpern J. P., 2005, ApJ, 633, 1114

\bibitem[\protect\citeauthoryear{Kaplan et al.}{2003}]{Kaplan03}
Kaplan D. L., Van Kerkwijk M. H., Marshall H. L., Jacoby B. A.,
Kulkarni S. R., 2003, ApJ, 590, 1008

\bibitem[\protect\citeauthoryear{Kargaltsev et al.}{2002}]{Kar02}
Kargaltsev O., Pavlov G. G., Sanwal D., Garmire G. P., 2002, ApJ,
580, 1060

\bibitem[\protect\citeauthoryear{Lipunov}{1992}]{lip92}
Lipunov V. M., 1992, Astrophysics of Neutron Stars, Springer-Verlag, Berlin

\bibitem[\protect\citeauthoryear{Lo Curto et al.}{2007}]{Lo07}
Lo Curto G., Mignani R. P., Perna R., Israel G. L., 2007, A\&A, 473,
539

\bibitem[\protect\citeauthoryear{Madsen}{1999}]{Madsen99}
Madsen J., 1999, in Hadrons in Dense Matter and Hadrosynthesis, 162, Springer, Berlin

\bibitem[\protect\citeauthoryear{Manchester et al.}{2005}]{Manchester05}
Manchester R. N., Hobbs G. B., Teoh A., Hobbs M., 2005, AJ, 129, 1993-2006

\bibitem[\protect\citeauthoryear{Mannarelli, Rajagopal \& Sharma}{2007}]{mrs07}
Mannarelli M., Rajagopal K., Sharma R. 2007, Phys. Rev. D76, 4026

\bibitem[\protect\citeauthoryear{Manzali, De Luca \& Caraveo}{2007}]{Manzali07}
Manzali A., De Luca A., Caraveo P. A., 2007, ApJ, 669, 570

\bibitem[\protect\citeauthoryear{McGowan et al.}{2003}]{McGowan03}
McGowan K. E., Kennea J. A., Zane S., C$\acute{\rm o}$rdova F. A.,
Cropper M., Ho C., Sasseen T., Vestrand W. T., 2003, ApJ, 591, 380

\bibitem[\protect\citeauthoryear{McGowan et al.}{2004}]{McGowan04}
McGowan K. E., Zane S., Cropper M., Kennea J. A., C$\acute{\rm
o}$rdova F. A., Ho C., Sasseen T., Vestrand W. T., 2004, ApJ, 600,
343

\bibitem[\protect\citeauthoryear{McGowan et al.}{2006}]{McGowan06}
McGowan K. E., Zane S., Cropper M., Vestrand W. T., Ho C., 2006,
ApJ, 639, 377

\bibitem[\protect\citeauthoryear{Mereghetti}{2008}]{Mere08}
Mereghetti S., 2008, ARAA, 15, 225

\bibitem[\protect\citeauthoryear{Michel}{1988}]{alpha}
Michel F. C., 1988, Phys. Rev. Lett., 60, 677

\bibitem[\protect\citeauthoryear{Owen}{2005}]{Owen05}
Owen B. J. 2005, Phys. Rev. Lett., 95, 211101

\bibitem[\protect\citeauthoryear{Pavlov \& Luna}{2009}]{luna09}
Pavlov G. G., Luna G. J. M., 2009, ApJ, 703, 910

\bibitem[\protect\citeauthoryear{Pavlov, Sanwal \& Teter}{2004}]{pav04}
Pavlov G. G., Sanwal D., Teter M. A., 2004, in: {\em Young Neutron
Stars and Their Environments}, IAU Symp. 218, Eds. F. Camilo \& B.
M. Gaensler, San Francisco (arXiv:0311526v1)

\bibitem[\protect\citeauthoryear{Pavlov et al.}{2002}]{PZST02}
Pavlov G. G., Zavlin V. E., Sanwal D., Tr$\ddot{\rm u}$mper, J.,
2002, ApJ, 569, L95

\bibitem[\protect\citeauthoryear{Pavlov, Kargaltsev \& Brisken}{2008}]{pav08}
Pavlov G. G., Kargaltsev O., Brisken W. F., 2008, ApJ, 675, 683

\bibitem[\protect\citeauthoryear{Possenti, Mereghetti \& Colpi}{1996}]{Possenti96}
Possenti A., Mereghetti S., Colpi M., 1996, A\&A, 313, 565

\bibitem[\protect\citeauthoryear{Schwope et al.}{2005}]{Schwope05}
Schwope A. D., Hambaryan V., Haberl F., Motch C., 2005, A\&A, 441,
597

\bibitem[\protect\citeauthoryear{Shuryak}{2009}]{Shuryak}
Shuryak E. V. 2009, Prog. Part. \& Nucl. Phys., 62, 48

\bibitem[\protect\citeauthoryear{Slane et al.}{2004a}]{SHVM04}
Slane P., Helfand D. J., Van de Swaluw E., Murry S. S., 2004a, ApJ, 616, 403

\bibitem[\protect\citeauthoryear{Thompson}{1997}]{Thomp97}
Thompson D. J., Harding A. K., Hermsen W., Ulmer M. P., 1997, in AIP
Conf. Proc. 410, Proc. Fourth Compton Symposium, ed. C. D. Dermer,
M. S. Strickman, J. D. Kurfess (New York: AIP), 39

\bibitem[\protect\citeauthoryear{Treves et al.}{2000}]{tre00}
Treves A., Turolla R., Zane S., Colpi M., 2000, PASP, 112, 297

\bibitem[\protect\citeauthoryear{Tsuruta}{2009}]{Tsuruta09}
Tsuruta S., 2009, Neutron Stars and Pulsars, Springer, Berlin

\bibitem[\protect\citeauthoryear{Usov}{2001}]{usov01}
Usov V. V., 2001, ApJ, 550, L179

\bibitem[\protect\citeauthoryear{Vonsovsky \& Katsnelson}{1989}]{VK89}
Vonsovsky S. V., Katsnelson M. I., 1989, Quantum Solid-State Physics, Springer-Verlag, Berlin

\bibitem[\protect\citeauthoryear{Wang et al.}{1998}]{WRHZ98}
Wang F. Y. -H., Ruderman M. A., Halpern J. P., Zhu T., 1998, ApJ,
498, 373

\bibitem[\protect\citeauthoryear{Weisskopf et al.}{2004}]{WOPEBTS04}
Weisskopf M. C., O'Dell S. L., Paerels F., Elsner R. F., Becker W.,
Tennant A. F., Swartz D. A., 2004, ApJ, 601, 1050

\bibitem[\protect\citeauthoryear{Xu \& Qiao}{1999}]{xq99}
Xu R. X., Qiao G. J., 1999, Chin. Phys. Lett., 16, 778

\bibitem[\protect\citeauthoryear{Xu}{2002}]{xu02}
Xu R. X., 2002, ApJ, 570, L65

\bibitem[\protect\citeauthoryear{Xu}{2003}]{xu03}
Xu R. X., 2003, ApJ, 596, L59

\bibitem[\protect\citeauthoryear{Xu}{2005}]{xu05}
Xu R. X., 2005, MNRAS, 356, 359

\bibitem[\protect\citeauthoryear{Xu}{2010}]{xu10}
Xu R. X., 2010, Int. Jour. Mod. Phys. D19, 1437

\bibitem[\protect\citeauthoryear{Xu}{2009}]{xu09}
Xu R. X., 2009, J. Phys. G: Nucl. Part. Phys. 36, 064010

\bibitem[\protect\citeauthoryear{Yakovlev et al.}{2008}]{Yak08}
Yakovlev D. G., Gnedin O. Y., Kaminker A. D., Potekhin A. Y., 2008,
AIPC, 983, 379

\bibitem[\protect\citeauthoryear{Zavlin}{2007a}]{Zav07a}
Zavlin V. E., 2007, ApJ, 665, L143

\bibitem[\protect\citeauthoryear{Zavlin \& Pavlov}{2004}]{ZP04}
Zavlin V. E., Pavlov G. G., 2004, ApJ, 616, 452

\bibitem[\protect\citeauthoryear{Zavlin, Tr$\ddot{\rm u}$mper \& Pavlov}{1999}]{ZTP99}
Zavlin V. E., Tr$\ddot{\rm u}$mper J., Pavlov G. G., 1999, ApJ, 525, 959

\bibitem[\protect\citeauthoryear{Zhang \& Harding}{2000}]{ZH00}
Zhang B., Harding A. K., 2000, ApJ, 532, 1150

\bibitem[\protect\citeauthoryear{Zhang, Xu \& Zhang}{2004}]{zxz04}
Zhang X., Xu R. X., Zhang S. N., 2004, in: {\em Young Neutron Stars
and Their Environments}, IAU Symp. 218, Eds. F. Camilo \& B. M.
Gaensler, San Francisco, pp.303

\bibitem[\protect\citeauthoryear{Zhu \& Xu}{2004}]{zhu04}
Zhu W. W., Xu R. X., 2004, preprint (arXiv:astro-ph/0410265)

\bibitem[\protect\citeauthoryear{Zhu et al.}{2009}]{zhu09}
Zhu W. W., Kaspi V. M., Gonzalez M. E., Lyne A. G., 2009, ApJ, in press (arXiv:0909.0962v1)

\end{thebibliography}
\end{document}